\documentstyle[mprocl]{article}

\def\Journal#1#2#3#4{{#1} {\bf #2}, #3 (#4)}

\def\NPB{{\em Nucl. Phys.} B}
\def\PLB{{\em Phys. Lett.}  B}

\def\PRD{{\em Phys. Rev.} D}

\def\al{\alpha}

\def\be{\begin{equation}}
\def\ee{\end{equation}}
\def\bea{\begin{eqnarray}}
\def\eea{\end{eqnarray}}

\begin{document}

\title{Black Hole in Higher Curvature Gravity 
        and AdS/CFT Correspondence}

\author{H. SAIDA}

\address{GSHES, Kyoto University, Kyoto,
 606-8501, Japan \\E-mail: saida@phys.h.kyoto-u.ac.jp}   

\author{J. SODA}

\address{Department of Fundamental Sciences, 
 FIHS, Kyoto, 606-8501
Japan \\E-mail: jiro@phys.h.kyoto-u.ac.jp}  


\maketitle
\abstracts{ 
The classical central charge for the higher curvature gravity in 3-dimensions
 is calculated using the Legendre transformation method. 
The statistical entropy of BTZ black hole is derived by the Cardy's formula
 and the result completely coincides with the Iyer-Wald formula for the 
 geometrical entropy. This coincidence suggests the generalized AdS/CFT
 correspondence. 
}

\section{Introduction}

It is now well established that there exists the correspondence
between the  Supergravity theory on Anti-deSitter spacetime (AdS) 
and the Large N limit of Super Yang-Mills CFT.~\cite{malda}
In the case of 3-d BTZ black hole ,  Strominger  has given a simple statistical
derivation of the  Bekenstein-Hawking entropy
 via Cardy's formula.~\cite{strom}
 This is an important check for the correspondence.

 The  AdS/CFT correspondence conjectured by Maldacena  
is the correspondence between the Type IIB Superstring theory
 on AdS  and the  Super Yang-Mills CFT.
The string theory in general receives corrections like as 
$  S_{\rm low \ energy}
     =R + \alpha' R^{\mu\nu\rho\sigma} R_{\mu\nu\rho\sigma} 
          +\cdots  \ .
$
Hence, a modest first step to the final aim is to extend 
 AdS/CFT correspondence to higher curvature gravity.
The geometrical entropy in the higher curvature gravity 
can be calculated by using the Noether charge method.~\cite{IW}
 The purpose of this paper is to give an evidence of AdS/CFT correspondence 
in higher curvature gravity by deriving the Iyer-Wald formula 
 in 3-dimensions using AdS/CFT correspondence.~\cite{saida}

\section{BTZ Black Hole Entropy in the Higher Curvature Gravity}

 Let us calculate the statistical entropy of the BTZ 
black hole in the higher curvature gravity. The action we treat here 
is of the form:
  $
    I = 1/16\pi G \int d^3 x \,
        f(R, R_{\mu \nu}) \sqrt{-g}  
  $
where $f(R, R_{\mu \nu})$ is the Lagrangian of general 
higher curvature gravity. This action is the most 
generic form of the three dimensional higher curvature gravity 
because the Weyl tensor vanishes in three dimensional spacetime. 
We restrict our treatment hereafter to the case that 
the spacetime is of the constant curvature. Further we assume 
that the curvature $R$ is of negative. Under such conditions, the 
BTZ black hole  can exist. 
The metric of the BTZ Black Hole is given by
\begin{equation}
    ds^2 = -(({r\over l})^2 +({4GJ\over r})^2 -{GM\over l} ) dt^2 
           + \frac{dr^2 }{({r\over l})^2 +({4GJ\over r})^2 -{GM\over l}}    
           + r^2 \left[ d\phi -{4GJ\over r^2} dt \right]^2 \nonumber 
\end{equation}
where $l$ is the curvature scale of the Anti-deSitter spacetime.
 Here mass $M$ and the angular momentum $J$ characterize the black hole.
The event horizon  can be read off from the metric as
$r_{+} = \sqrt{ 2Gl(M+J) }  + \sqrt{ 2Gl(M-J) }$. 
The geometrical entropy of the black hole can be 
calculated  to be
\begin{eqnarray}
    S_{IW}
      = - {1 \over 8G} \oint_{H} d\phi \sqrt{g_{\phi\phi}} \,
          { \partial f \over \partial R_{\mu\al} } \, g^{\nu\beta} \,
          \epsilon_{\mu\nu}\epsilon_{\al\beta} 
      =   \frac{1}{12G} \, g^{\mu\nu}
          \frac{\partial f}{\partial R_{\mu\nu}} \,
          \oint_{r_+} d\phi \sqrt{g_{\phi\phi}} 
\end{eqnarray}
where $H$, $h$ and $\epsilon_{\mu\nu}$ are the spatial section of the 
event horizon, the determinant of the induced metric on $H$, and  
   the binormal to $H$, respectively. 

In order to calculate the central charge for this theory,
  Legendre transformation~\cite{mag} 
$
    \bar{g}^{\mu\nu} \equiv
      \left[
        -\det \left( {\partial (f \sqrt{-g}) }/
{\partial R_{\alpha\beta}}
      \right) \right]^{-1}
      {\partial (f \sqrt{-g}) }/{\partial R_{\mu\nu}} 
$
is useful. 
The point is that the action can be rewritten as 
the Einstein-Hilbert action of $\bar{g}^{\mu\nu}$ and 
an auxiliary matter field. 
Because the spacetime of the BTZ black hole is of constant 
curvature, substituting
 $
    \bar{g}_{\mu\nu} = \Omega^{2} g_{\mu\nu} 
 $
into the Legendre transformation formula gives
  $
    \Omega =  \, g^{\mu\nu} \partial f /\partial R_{\mu\nu}/3 
  $.
 The calculation of the central charge in the Einstein frame can be 
carried out by relating the quantities in the Einstein frame to 
those in the original frame through the conformal transformation.
The central charge is calculated to be
$
    c = {l}/{2G} \, g^{\mu\nu}
          {\partial f}/{\partial R_{\mu\nu}} 
$
Although this is the central charge calculated in the Einstein 
frame, this central charge is equivalent to that in the original 
higher curvature frame.

 The mass and the angular momentum in 
the higher curvature gravity are calculated through the Noether 
charge form  to be $\Omega M$ and 
$\Omega J$, respectively. Then the two Virasoro eigen values 
$\lambda$ and 
$\tilde{\lambda}$ in the higher curvature gravity are given by,
$
    \lambda = \Omega \, (M+J) /2 \  , \ 
    \tilde{\lambda} = \Omega \, (M-J) /2 \, .
$
Finally the statistical entropy of the BTZ black hole in the higher 
curvature gravity can be deduced from Cardy's formula~\cite{cardy}
\begin{equation}
    S_C
     = 2\pi \sqrt{ \frac{c \, \lambda}{6} }
              + 2\pi \sqrt{ \frac{c \, \tilde{\lambda} }{6} } 
     = \frac{\pi}{12G} \, g^{\mu\nu}
          \frac{\partial f}{\partial R_{\mu\nu}} \,
          \left[ \, \sqrt{ 8Gl(M+J) }
                + \sqrt{ 8Gl(M-J) } \, \right]  
\end{equation}
This formula 
 coincides with the statistical entropy $S_{IW} $.~\cite{saida}
\section{Summary }
We have shown  
$ S_{\rm IW} = S_{\rm statistical} $ 
in 3-dimensional higher curvature gravity. 
{\bf This is an evidence of AdS/CFT correspondence in higher curvature
gravity.} \\
This also suggests the string-CFT correspondence.
%

%

\eject

\end{document}